\documentclass[12pt,preprint]{aastex}

\begin{document}

\title{\bf Mira~B Rejuvenated\altaffilmark{1}}

\author{Brian E. Wood\altaffilmark{2} and Margarita Karovska\altaffilmark{3}}

\altaffiltext{1}{Based on observations with the NASA/ESA Hubble Space
  Telescope and the NASA-CNES-CSA Far Ultraviolet Spectroscopic Explorer.
  The former observations were obtained at the Space Telescope Science
  Institute, which is operated by the Association of Universities for
  Research in Astronomy, Inc., under NASA contract NAS5-26555.
  FUSE is operated for NASA by the Johns Hopkins University under NASA
  contract NAS5-32985.}
\altaffiltext{2}{JILA, University of Colorado, Boulder, CO
  80309-0440; woodb@origins.colorado.edu.}
\altaffiltext{3}{Smithsonian Astrophysical Observatory, 60 Garden St., 
  Cambridge, MA 02138; mkarovska@cfa.harvard.edu.}

\begin{abstract}

     Recent ultraviolet spectra from the {\em Hubble Space Telescope}
(HST) and {\em Far Ultraviolet Spectroscopic Explorer} (FUSE) satellites
demonstrate that UV line and continuum fluxes observed from Mira~B are
increasing back towards the levels that the {\em International Ultraviolet
Explorer} observed in 1979--1980 and 1990--1995, after having been
found in a very low state by HST and FUSE in 1999--2001.  The UV emission
is associated with accretion of material onto Mira~B from Mira~A's
massive wind, so the variability is presumably due to variations
in accretion rate.  From wind absorption features, we estimate a
Mira~B mass loss rate of $2.5\times 10^{-12}$ M$_{\odot}$~yr$^{-1}$,
indicating that Mira~B's wind has increased in strength along with the
accretion rate.  The UV variability may be associated with a previously
reported 14-year periodicity in Mira~B's optical emission.

\end{abstract}

\keywords{accretion, accretion disks --- binaries: close --- stars:
  individual (o Ceti) --- stars: winds, outflows --- ultraviolet: stars}

\section{INTRODUCTION}

     Mira~AB is the nearest symbiotic system at a distance of only
$\sim130$~pc, \citep{macp97}, and the only symbiotic binary in
which the components of the system have been resolved at wavelengths
ranging from X-rays to radio \citep{mk97,mk05,ldm06}.  The primary (Mira,
o~Ceti, HD~14386) is a very large pulsating AGB star (photospheric
diameter of $\sim500$~R$_{\odot}$), with a massive, cool wind of
$\sim 10^{-7}$ M$_{\odot}$~yr$^{-1}$ \citep{pfb88}.
The companion (Mira~B, VZ~Ceti) is accreting material from Mira~A, but
the exact nature of the accretor is uncertain \citep{mj84};
it has usually been assumed to be a white dwarf \citep{ahj54,mk05}.
The current angular separation between the components is
$\sim$ $0.6^{\prime\prime}$ \citep{mk97}.  The orbit of the
binary is poorly known, but \citet{jlp02} estimate a long
orbital period of $\sim 500$ years, and this orbit suggests that the
distance from Mira~A to Mira~B is close to the apparent separation in
the plane of the sky, about 70~AU.  This interacting binary, being one
of the few accretion systems in which the components of the system are
resolvable, provides a unique opportunity to study the spectral energy
distributions of the components of an interacting binary individually,
as well as being a laboratory for testing accretion models.

     In 1995, the Faint Object Camera (FOC) instrument on the {\em Hubble
Space Telescope} (HST) resolved the Mira system for the first time
spatially and spectrally at UV and optical wavelengths \citep{mk97}.
Since then, the system has been monitored from UV to radio
wavelengths.  In the past few years, we have been witnessing dramatic
changes in Mira~B, especially at UV wavelengths \citep{mk97,bew01}.
Spectra from the Space Telescope Imaging
Spectrograph (STIS) instrument on HST imply a substantial decrease in
accretion luminosity in the late 1990s, compared with fluxes observed at
earlier times by the {\em International Ultraviolet Explorer} (IUE)
and HST/FOC.  Furthermore, a forest of Ly$\alpha$-fluoresced H$_{2}$
emission lines dominated the HST spectra of Mira~B in 1999, despite not
being seen at all in any previous IUE or HST spectra \citep{bew01,bew02}.
Observations from the {\em Far Ultraviolet Spectroscopic
Explorer} (FUSE) demonstrated that the low accretion luminosity and
prominent H$_{2}$ emission persisted into 2001 \citep{bew04}.
On 2003 December 6, {\em Chandra} observations detected an  unprecedented
soft X-ray outburst from Mira~A \citep{mk05}, further
evidence of remarkable dynamicism in the Mira~AB system.  The outburst
was also detected in contemporaneous H$\alpha$ observations and is
likely associated with a magnetic flare followed by a mass ejection or
jet-like activity \citep[][Karovska 2006, in prep.]{mk05}.

     We here present recent UV observations of Mira~B from HST and FUSE,
obtained following the X-ray outburst. We compare these data with past
observations by these satellites and by IUE.  We discuss the long term
variations of Mira~B and a possible correlation with long term
optical variability.

\section{UV SPECTROSCOPY OF MIRA B FROM 1979--2001}

     Numerous UV spectra of Mira~AB were obtained in 1979--1980 and
1990--1995 by IUE.  This data set includes 32 spectra taken in IUE's
short-wavelength, low-resolution mode (i.e., SW-LO spectra,
covering $\sim 1150-1950$~\AA).  These far-UV (FUV) spectra are dominated
by broad, hot lines such as C~II $\lambda$1335, Si~IV $\lambda$1400,
C~IV $\lambda$1550, Si~III] $\lambda$1892, and C~III] $\lambda$1908,
which are produced by material accreting onto Mira~B from
Mira~A's massive wind.  The few high-resolution spectra of this
wavelength region (i.e., SW-HI spectra) are very noisy, but they suggest
very broad widths for these lines, consistent with emission from a
rapidly rotating accretion disk \citep{dr85}.  Unlike
IUE, HST can resolve the binary, and UV spectra from HST confirm
that the broad lines are from Mira~B, and that Mira~A contributes very
little to the IUE spectra below 2000~\AA\ \citep{mk97,bew01}.

     Only a modest level of UV variability is present within the large IUE
SW-LO data set, with line and continuum fluxes varying by about
a factor of 2.  However, when HST observed Mira~B in 1999 August using
STIS, the character of Mira~B's FUV spectrum had changed dramatically.
The UV continuum and all of the aforementioned UV emission lines had
weakened by at least a factor of 20, and the FUV spectrum was now dominated
by many narrow H$_{2}$ lines fluoresced by H~I Ly$\alpha$ \citep{bew01},
which are not detectable at all in the IUE spectra.
The accretion onto Mira~B results in a warm, fast wind from the star
that is most clearly apparent in broad absorption features in the Mg~II
h \& k lines at 2800~\AA.  Given the presence of accretion, this wind
is envisioned to be of a bipolar nature.  Analysis of the Mg~II absorption
demonstrates that the substantial drop in UV flux from Mira~B seen by HST
is accompanied by a corresponding drop in its mass loss rate \citep{bew02}.

     The low UV fluxes observed by HST in 1999 are most naturally
explained by a substantial decrease in the accretion rate onto Mira~B,
which also led to a weaker wind from the star.  The sudden prominence
of the H$_{2}$ lines in the HST spectra suggests that unlike other UV
lines, the H~I Ly$\alpha$ line responsible for fluorescing these lines
had {\em not} decreased in strength relative to the IUE era.  Studying
Ly$\alpha$ in IUE spectra is very difficult due to geocoronal
contamination, but \citet{bew02} still uncovered evidence to
support the contention that Ly$\alpha$ fluxes are similar in the HST
and IUE spectra.

     A 2001 November observation from FUSE found Mira~B to be still in a
very low state, with C~III $\lambda$1175 fluxes comparable to those in
the 1999 HST/STIS spectrum \citep{bew04}.  Like the STIS data,
this FUSE spectrum also shows numerous narrow H$_{2}$ lines.
Ground-based observations have suggested that Mira~B's optical
light curve may have a period of 14 years, with the last published minimum
being in 1971 \citep{ahj54,yy77}. \citet{bew04}
noted that the the numerous IUE SW-LO observations are all
near predicted maxima of this cycle, while the 1999 HST and 2001 FUSE
observations are near a predicted minimum.  Thus, it is possible that
the observed long-term UV variability could be correlated with previously
detected optical variations. If the optical periodicity is real and if a
connection between the optical and UV light curves truly exists, Mira~B's
UV fluxes should increase towards a 2006--2007 maximum predicted by the
14-year optical periodicity.  The new 2004 HST and FUSE spectra presented
here allow us to test this prediction, and to see
if the character of Mira~B's UV spectrum has changed once again.

\section{RECENT HST AND FUSE OBSERVATIONS}

\subsection{Spectral Line Variability}

     Mira~B was observed by HST with the STIS instrument on 2004
February 16.  A near-UV (NUV) spectrum from 2274--3119~\AA\ was taken
using the moderate resolution E230M grating, with an exposure time
of 600~s, and Mira~B's FUV spectrum from 1140--1729~\AA\ was observed
using the E140M grating, with an exposure time of 2100~s.  Both spectra were
taken through the $0.2^{\prime\prime}\times 0.2^{\prime\prime}$ aperture.

     Mira was also observed by FUSE on 2004 August 31.  This was an
8603~s exposure through the LWRS aperture, covering the
905--1187~\AA\ spectral range.  This represents only a fraction of the
requested 30~ksec exposure, so the new FUSE spectrum is much noisier
than the 2001 FUSE observation \citep{bew04}.  A
C~III $\lambda$1175 flux can still be measured from the data and some
of the strongest H$_{2}$ lines can still be discerned.

     Figures~1 and 2 compare UV spectral regions observed by HST in 2004
with the old 1999 spectra.  Five of the Figure~1 panels focus on
spectral regions containing FUV emission lines of H~I, O~I, C~II, Si~IV,
and C~IV.  Some of these lines are blended with H$_{2}$ emission, so the
locations of strong, narrow H$_{2}$ lines are also indicated in the figure.
The last panel shows the NUV spectrum in the region of the Fe~II UV1
multiplet lines.  The line locations indicated in Figure~1 are with
respect to Mira's rest frame, where the rest wavelengths are taken from
\citet{ha93} and \citet{dcm03}, and we assume a radial velocity
of 56 km~s$^{-1}$ for the Mira system \citep{pfb88,pp90,ej00}.

     Figures~1 and 2 show that UV fluxes from Mira~B have increased
dramatically since 1999, with the O~I, C~II, Si~IV, C~IV, and Mg~II lines
being much stronger than they were.  The notable exceptions to this
rule, the H~I Ly$\alpha$ line and the H$_{2}$ lines fluoresced by
Ly$\alpha$, will be discussed further in \S3.3.  The UV continuum is also
much stronger in 2004 (see the Fe~II panel in Fig.~1).  In Figure~3, we
plot measured fluxes as a function of time for many different lines,
including measurements from IUE, HST, and FUSE.  For
C~III $\lambda$1175, O~I $\lambda$1300, C~II $\lambda$1335,
Si~IV $\lambda$1400, and C~IV $\lambda$1550, the IUE fluxes are measured
from the 32 IUE SW-LO observations mentioned in \S2.  Examples
of these spectra have been previously presented by \citet{dr85}
and \citet{bew01}.  Note that the fluxes in Figure~3 are
the total flux of these multiplets (e.g., for Si~IV the flux includes
both the $\lambda$1393 and $\lambda$1402 lines), given that the
individual components of the multiplets are not resolved in the IUE SW-LO
spectra.

     Blended H$_{2}$ lines are removed from
the data before measuring fluxes from the HST and FUSE data.  This
cannot be done for the low resolution IUE spectra, but thanks to the
higher line fluxes seen by IUE, H$_{2}$ is not as strong a contaminant
in the IUE data, so the correction is not as important \citep{bew01}.
The C~III $\lambda$1175 line is the only line observable by
all three satellites, though in the HST/STIS spectra and the 2004 FUSE
spectrum the C~III line is observed at very low S/N.  In the higher quality
2001 FUSE spectrum, the H$_{2}$ lines add
$1.6\times 10^{-14}$ ergs~cm$^{-2}$~s$^{-1}$ to the C~III feature
\citep{bew04}, so we simply subtract this flux from each C~III
flux measurement to correct for the strong H$_{2}$ contamination in C~III.

     The IUE's inability to resolve the Mira binary is generally not
a problem since Mira~A does not contribute significantly to most UV
emission lines.  However, Mira~A can contribute to Mira~AB's Mg~II lines
at certain pulsation phases \citep{bew00}.  Fortunately, in
long-wavelength, high-resolution (LW-HI) IUE spectra, Mira~A's
narrow Mg~II lines are mostly contained within the saturated core of
Mira~B's Mg~II wind absorption.  Therefore, the Mira~A Mg~II line can
be removed before measuring a Mg~II flux for Mira~B.  The IUE Mg~II
fluxes plotted in Figure~3 are all from LW-HI spectra.  We do not use
LW-LO spectra because of potential Mira~A contamination.  The Mg~II
fluxes in Figure~3 include both the h and k lines.

     Figure~3 shows the dramatic decrease in UV line fluxes
observed in 1999--2001.  The highest temperature line, C~IV $\lambda$1550, 
exhibits a particularly large decrease of nearly a factor of 100.
The fluxes increase significantly in 2004, but they are still lower
than fluxes observed by IUE in 1979--1980 and 1990--1995.
As discussed in \S2, a 14-year periodicity
has been detected in Mira~B's optical light curve \citep{ahj54,yy77},
and dotted lines in Figure~3 indicate the
phase and period of this light curve.  The increase in UV flux that we
see in our 2004 spectra is roughly consistent with the predictions of the
optical light curve, assuming that the UV emission follows the same
periodic behavior as that inferred for the optical light.
It therefore remains a plausible hypothesis that Mira~B's long-term
UV variability is associated with the previously detected periodic
optical variability.  However, truly verifying this connection would
require UV observations over at least another cycle.  

\subsection{Wind Absorption}

     The blue sides of all the resonance lines in Figures~1 and 2 are
suppressed by absorption from Mira~B's wind, with the exception of C~IV
(see below).  The wind absorption is more readily apparent in the new
2004 data than in the old 1999 spectra.  This is in part due to much
better S/N in the new data thanks to the higher
fluxes, but it is also due to the wind absorption being intrinsically
stronger, which is particularly evident for the Fe~II and Mg~II lines.
The mass loss rate of Mira~B's wind can be estimated
most easily from the Mg~II wind absorption profiles, which have the
highest S/N.

     From analyses of the Mg~II k line at 2796~\AA, we have previously
estimated a mass loss rate of
$\dot{M}=5\times 10^{-13}$ M$_{\odot}$ yr$^{-1}$ from the 1999 HST spectrum,
and $\dot{M}=1\times 10^{-11}$ M$_{\odot}$ yr$^{-1}$ from
a typical IUE spectrum \citep{bew02}.  Using the exact
same methodology, we have fitted a wind
absorption profile to the 2004 Mg~II k line in Figure~2.  The
observed absorption profile is not smooth, suggesting a more complicated
and structured wind than we can fit with such a simple model.
Nevertheless, the mass loss rate inferred by the fit should still be
a reasonable estimate.  This fit suggests
a mass loss rate of $\dot{M}=2.5\times 10^{-12}$ M$_{\odot}$ yr$^{-1}$
and a wind terminal velocity of $V_{\infty}=450$ km~s$^{-1}$.
For comparison, the absorption predicted by the wind absorption profile
from 1999 is also shown in Figure~2,
assuming $\dot{M}=5\times 10^{-13}$ M$_{\odot}$ yr$^{-1}$ and
$V_{\infty}=250$ km~s$^{-1}$.  The wind absorption is significantly
stronger in 2004, meaning that Mira~B's wind has clearly
strengthened, but the mass loss rate is still not as high as the
$\dot{M}=1\times 10^{-11}$ M$_{\odot}$ yr$^{-1}$ value estimated for
the IUE era.  The mass loss rate variations are consistent with
the accretion rate variations inferred from
the UV fluxes (see Fig.~3).  Clearly, Mira~B's accretion rate is highly
correlated with the strength of its wind.

     The only atomic lines in Figure~1 that do not definitively possess
wind absorption are the C~IV lines.  Small decreases in flux blueward of
the center of the two C~IV lines may indicate the presence of {\em some}
C~IV wind absorption, but this absorption is not nearly as prominent as
the saturated absorption seen for the other lines.  The presence of
strong wind absorption for the Si~IV $\lambda$1392 line, and the
relative lack of absorption for C~IV
implies a maximum temperature for the wind between $T=6.0\times 10^{4}$~K
(Si~IV's line formation temperature) and $T=1.0\times 10^{5}$~K (C~IV's
line formation temperature).

\subsection{The Anomalous Ly$\alpha$ Line}

     The one clearly anomalous line in Figure~1 is H~I Ly$\alpha$.
Although the 2004 Ly$\alpha$ flux is slightly higher than in 1999, this
line does not show the dramatic flux increase that the other emission
lines show.  The singular lack of variability for this line was also
noted by \citet{bew02}, who inferred that the Ly$\alpha$ flux during
the IUE era cannot have been much stronger than in 1999, unlike all the
other lines.  The lack of variability for Ly$\alpha$ means that the H$_{2}$
lines that are fluoresced by Ly$\alpha$ are also relatively constant,
as shown by the H$_{2}$ lines blended with C~II and Si~IV in Figure~1.

     It is difficult to understand why the Ly$\alpha$ line behaves
so differently from the continuum and all other emission lines, including
lines formed at similar temperatures (O~I, C~II, and Mg~II).
Does this mean that the Ly$\alpha$ emission is coming from a
different place than the other lines?  If so, why?  Another possibility
is that the Ly$\alpha$ flux actually {\em does} vary along with the
other lines, but the extremely high opacity of Mira~B's wind
to Ly$\alpha$ moderates the variation.  Wind absorption explains why
the observed Ly$\alpha$ emission is entirely redshifted relative to
line center (see Fig.~1).  Analogous absorption is also seen for many of
the other lines, but wind opacity will be orders of magnitude higher for
Ly$\alpha$ since hydrogen is much more abundant than other atomic species.

     Given that increases in UV flux are clearly accompanied by increases
in wind strength, perhaps the enhanced wind obscuration largely cancels the
flux increase for Ly$\alpha$.  This hypothesis would require that Ly$\alpha$
photons are actually destroyed rather than just scattered by the wind.
Perhaps this destruction could be accomplished by scattering into Mira~B's
accretion disk, or extinction by dust entrained within Mira~B's wind.

\section{CAUSES OF LONG-TERM VARIABILITY}

     The fundamental cause of Mira~B's substantial variability
remains uncertain.  Variable mass loss from Mira~A could have caused
the changes in the UV flux of Mira~B observed in the past 10 years,
signaling that the system is undergoing important transformations.
These transformations may have set the stage for the remarkable X-ray
outburst from Mira~A that was observed by {\em Chandra} on
2003 December 6 \citep{mk05}.
%This outburst was not present in an XMM observation from
%2003 July 23 \citep{jhk04}, and had faded by the time of a
%follow-up {\em Chandra} observation on 2004 January 11.

     If the variability is associated with the possible 14-year optical
periodicity, the reason for the periodicity is a mystery.  The UV lines
in particular are believed to originate from accretion processes on Mira~B,
so a periodicity in these lines would suggest a periodicity for the accretion
rate.  Accretion instabilities could easily lead to variability, but it
is not clear why they would be periodic.  Perhaps the gravitational
influence of an undetected stellar or planetary companion could lead to
a corresponding periodicity in the accretion rate, perhaps through
wind focusing effects.  Hints of a third
body in the Mira~AB system have been previously reported in the literature
\citep{pb80,mk92,mk93}.  Since Mira~A's wind is the source of the
accreting material, a periodicity in Mira~A's mass loss rate could in
principle also lead to periodicity in the accretion rate.
No such periodicity has yet been found in the light curve of Mira
compiled by AAVSO, but searches for very low amplitude, long-term
periodicity are ongoing (M.\ Templeton, private communication).

     We hope to explore these possibilities further in the future,
as we reach the expected maximum in 2006--2007 and over the next
14-year cycle, using different observational techniques including
optical/IR interferometry.  However, future UV high resolution spectra
are very much in doubt due to the failure of the STIS instrument in
2004 August, and the reaction wheel and gyro failures on FUSE that
currently limit observations to high declinations.  Thus, new UV
instrumentation may be necessary to continue spectroscopic observations
in the ultraviolet.

\acknowledgments

This work was supported by NASA grant NAS8-39073 to the
Smithsonian Astrophysical Observatory and NASA grant NNG05GC14G to
the University of Colorado.
MK is a member of the Chandra X-ray Center, which is operated
by the Smithsonian Astrophysical Observatory under contract to 
NASA NAS8-39073.

\clearpage

\begin{figure}
\plotone{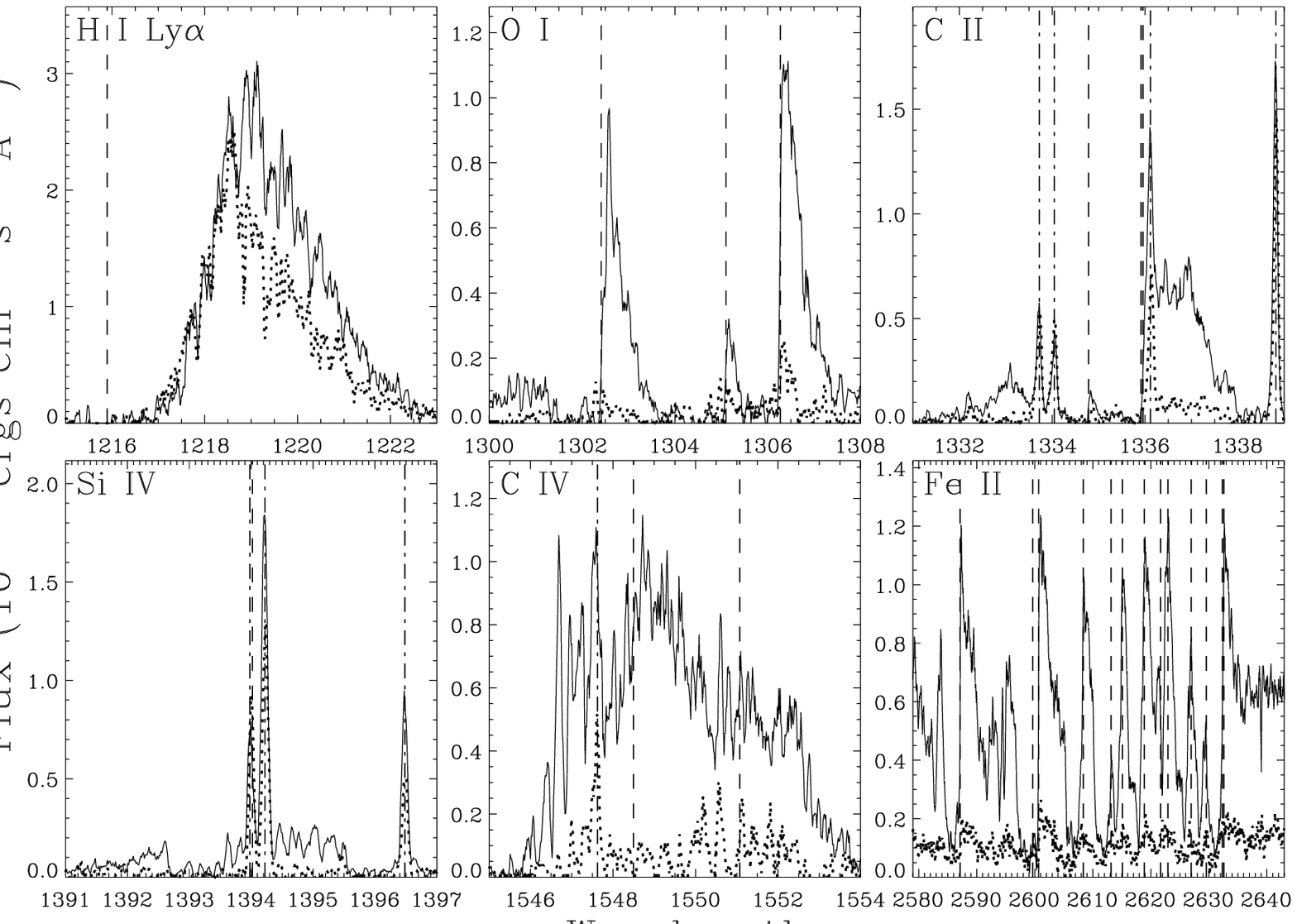}
\caption{A comparison of UV spectra of Mira~B taken in 1999 (dotted line)
  and 2004 (solid line).  The spectra are smoothed slightly for the sake
  of appearance.  Dashed lines indicate the rest wavelengths
  (in Mira's rest frame) of the line or lines indicated by the panel
  labels (e.g., H~I Ly$\alpha$, O~I, C~II, Si~IV, C~IV, and Fe~II).
  Dot-dashed lines indicate the locations of strong, narrow H$_{2}$
  lines.}
\end{figure}

\clearpage

\begin{figure}
\plotone{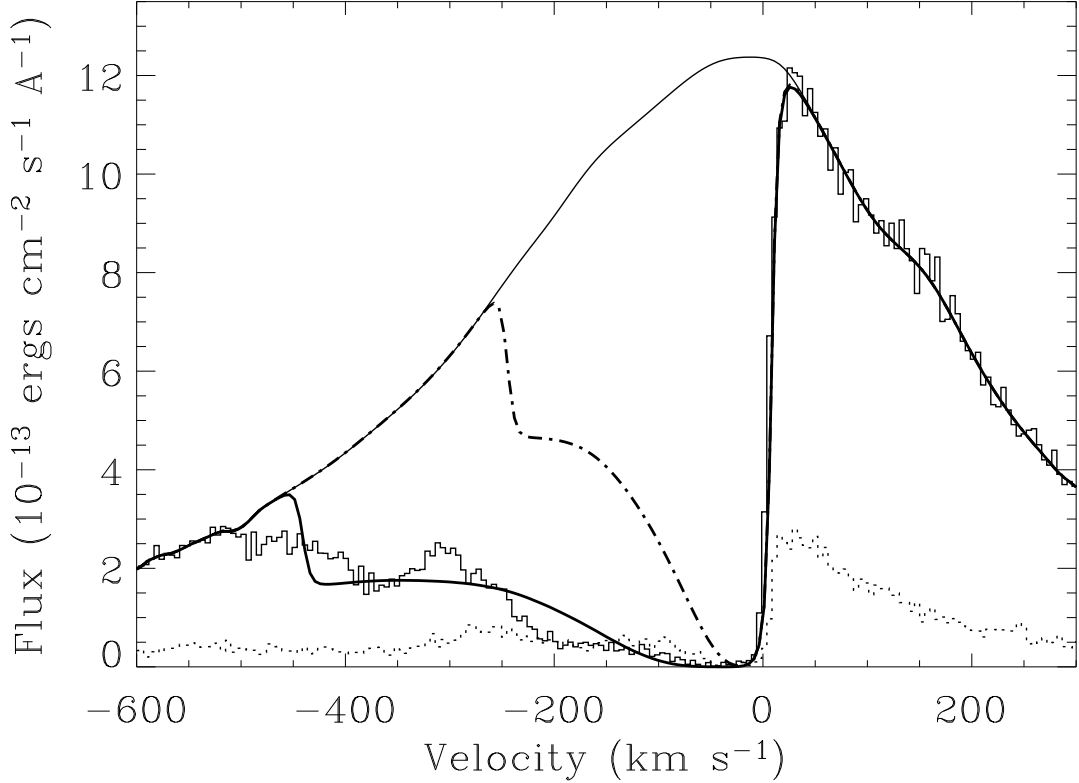}
\caption{A comparison of Mira~B's Mg~II k line profile from 1999 (dotted
  histogram) and 2004 (solid histogram), plotted on a velocity scale
  in Mira~B's rest frame.  A wind absorption profile has
  been fitted to the 2004 data (thick solid line), which suggests a mass
  loss rate of $\dot{M}=2.5\times 10^{-12}$ M$_{\odot}$ yr$^{-1}$ and a
  terminal velocity of $V_{\infty}=450$ km~s$^{-1}$.  Also shown is
  the absorption predicted by a fit to the 1999 wind absorption
  (dot-dashed line), assuming
  $\dot{M}=5\times 10^{-13}$ M$_{\odot}$ yr$^{-1}$ and
  $V_{\infty}=250$ km~s$^{-1}$ (Wood \& Karovska 2004).  This greatly
  underpredicts the amount of absorption observed in 2004, illustrating
  the increase in wind strength from 1999 to 2004.}
\end{figure}

\clearpage

\begin{figure}
\plotone{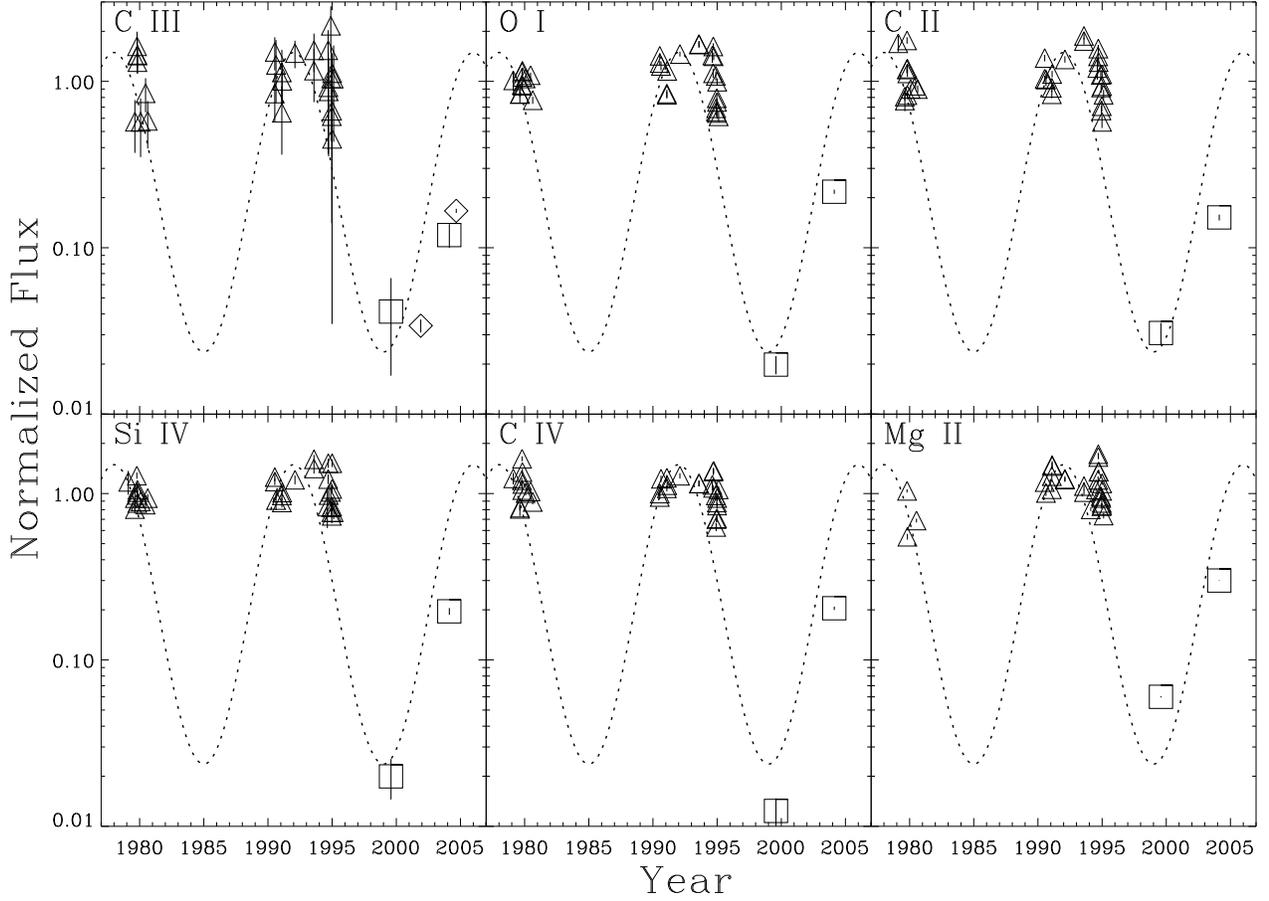}
\caption{Fluxes of various UV emission lines as a function of time, based
  on spectra from IUE (triangles), HST (boxes), and FUSE (diamonds).
  The fluxes are normalized to the average value observed by IUE.
  The lines and normalization factors (in units of
  ergs~cm$^{-2}$~s$^{-1}$) are
  C~III $\lambda$1175:  $4.1\times 10^{-13}$,
  O~I $\lambda$1300:  $8.1\times 10^{-13}$,
  C~II $\lambda$1335:  $6.5\times 10^{-13}$,
  Si~IV $\lambda$1400:  $5.5\times 10^{-13}$,
  C~IV $\lambda$1550:  $1.7\times 10^{-12}$, and
  Mg~II $\lambda$2800:  $1.3\times 10^{-11}$.
  The dotted lines are schematic representations of the 14 year
  periodicity of Mira~B suggested by optical observations (Joy 1954;
  Yamashita \& Maehara 1977).}
\end{figure}


\clearpage

\begin{thebibliography}{}

\bibitem[Abgrall et al.(1993)]{ha93}
Abgrall, H. A., Roueff, E., Launay, F., Roncin, J. -Y., \& Subtil, J. -L.
  1993, A\&AS, 101, 273
\bibitem[Baize(1980)]{pb80}
Baize, P. 1980, A\&AS, 39, 86
\bibitem[Bowers \& Knapp(1988)]{pfb88}
Bowers, P. F., \& Knapp, G. R. 1988, ApJ, 332, 299
\bibitem[Josselin et al.(2000)]{ej00}
Josselin, E., Mauron, N., Planesas, P., \& Bachiller, R. 2000, A\&A, 362, 255
\bibitem[Joy(1954)]{ahj54}
Joy, A. H. 1954, ApJS, 1, 39
\bibitem[Jura \& Helfand(1984)]{mj84}
Jura, M., \& Helfand, D. J. 1984, ApJ, 287, 785
\bibitem[Karovska(1992)]{mk92}
Karovska, M. 1992, in Complementary Approaches to Double and Multiple Star
  Research, ed. H. A. McAlister \& W. I. Hartkopf (ASP Conf. Ser., Vol. 32),
  558
\bibitem[Karovska et al.(1997)]{mk97}
Karovska, M., Hack, W., Raymond, J., \& Guinan, E. 1997, ApJ, 482, L175
\bibitem[Karovska et al.(1993)]{mk93}
Karovska, M., Nisenson, P. \& Beletic, J. 1993, ApJ, 402, 311
\bibitem[Karovska et al.(2005)]{mk05}
Karovska, M., Schlegel, E., Hack, W., Raymond, J. C., \& Wood, B. E. 2005,
  ApJ, 623, L137
%\bibitem[Kastner \& Soker(2004)]{jhk04}
%Kastner, J. H., \& Soker, N. 2004, ApJ, 616, 1188
\bibitem[Matthews \& Karovska(2006)]{ldm06}
Matthews, L. D., \& Karovska, M. 2006, ApJ, 637, L49
\bibitem[Morton(2003)]{dcm03}
Morton, D. C. 2003, ApJS, 149, 205
\bibitem[Perryman et al.(1997)]{macp97}
Perryman, M. A. C., et al. 1997, A\&A, 323, L49
\bibitem[Planesas et al.(1990)]{pp90}
Planesas, P., Bachiller, R., Martin-Pintado, J., \& Bujarrabal, V. 1990,
  ApJ, 351, 263
\bibitem[Prieur et al.(2002)]{jlp02}
Prieur, J. L., Aristidi, E., Lopez, B., Scardia, M., Mignard, F., \&
  Carbillet, M. 2002, ApJS, 139, 249
\bibitem[Reimers \& Cassatella(1985)]{dr85}
Reimers, D., \& Cassatella, A. 1985, ApJ, 297, 275
\bibitem[Wood \& Karovska(2000)]{bew00}
Wood, B. E., \& Karovska, M. 2000, ApJ, 535, 304
\bibitem[Wood \& Karovska(2004)]{bew04}
Wood, B. E., \& Karovska, M. 2004, ApJ, 601, 502
\bibitem[Wood et al.(2001)]{bew01}
Wood, B. E., Karovska, M., \& Hack, W. 2001, ApJ, 556, L51
\bibitem[Wood et al.(2002)]{bew02}
Wood, B. E., Karovska, M., \& Raymond, J. C. 2002, ApJ, 575, 1057
\bibitem[Yamashita \& Maehara(1977)]{yy77}
Yamashita, Y., \& Maehara, H. 1977, PASJ, 29, 319

\end{thebibliography}
\end{document}